\documentclass[reprint,amsmath,amssymb,aps,pre]{revtex4-1} 	
\pdfoutput=1

\usepackage{bmpsize}
\usepackage{graphicx,color}								
\usepackage{amssymb}
\usepackage{natbib}
\usepackage{array}
\usepackage{subfigure}
\usepackage[usenames,dvipsnames,svgnames,table]{xcolor}
\usepackage{epstopdf}

\newcommand{\paddyspeaks}[1]{{\color{black} #1}}

\newcommand{\maxspeaks}[1]{{\color{black} #1}}

\begin{document}

\title{Oil-in-water microfluidics on the colloidal scale: new routes to self-assembly and glassy packings}

\author{Max Meissner}
\affiliation{H.H. Wills Physics Laboratory, University of Bristol, Bristol, BS8 1TL, UK} 
\affiliation{Centre for Nanoscience and Quantum Information, Tyndall Avenue, Bristol, BS8 1FD, UK}

\author{Jun Dong}
\affiliation{H.H. Wills Physics Laboratory, University of Bristol, BS8 1TL, Bristol UK}
\affiliation{Centre for Nanoscience and Quantum Information, Tyndall Avenue, Bristol, BS8 1FD, UK}

\author{Jens Eggers}
\affiliation{Mathematics Department,University of Bristol, BS8 1TW, Bristol UK }

\author{Annela M. Seddon}
\affiliation{H.H. Wills Physics Laboratory, University of Bristol, BS8 1TL, Bristol UK}
\affiliation{Bristol Centre for Functional Nanomaterials, University of Bristol, Bristol, BS8 1TL, UK}
\affiliation{Centre for Nanoscience and Quantum Information, Tyndall Avenue, Bristol, BS8 1FD, UK}

\author{C. Patrick Royall}
\affiliation{H.H. Wills Physics Laboratory, University of Bristol, Bristol, BS8 1TL, UK} 
\affiliation{Chemistry Department, University of Bristol, Bristol, BS8 1TS, UK}
\affiliation{Department of Chemical Engineering, Kyoto University, Kyoto 615-8510, Japan}
\affiliation{Centre for Nanoscience and Quantum Information, Tyndall Avenue, Bristol, BS8 1FD, UK}

\begin{abstract}

\maxspeaks{We have developed Norland Optical Adhesive (NOA) flow focusing devices, making use of the excellent solvent compatibility and surface properties of NOA to generate micron scale oil-in-water emulsions with polydispersities as low as 5\%. While current work on microfluidic oil-in-water emulsification largely concerns the production of droplets with sizes on the order of 10s of micrometres, large enough that Brownian motion is negligible, our NOA devices can produce droplets with radii ranging from $2 \mu m$ to $12 \mu m$. To demonstrate the utility of these emulsions as colloidal model systems we produce fluorescently labelled polydimethylsiloxane droplets suitable for particle resolved studies with confocal microscopy. We analyse the structure of the resulting emulsion in 3D using coordinate tracking and the topological cluster classification and reveal a new mono-disperse thermal system.}

\end{abstract}

\maketitle

\section{Introduction} \label{Intro}
The \paddyspeaks{vast} majority of work on colloidal systems concerns itself either with the study of solid particles, for example solid colloidal spheres, or with emulsions, where an oil phase is dispersed in an aqueous medium or vice versa\cite{hunter2001}. For solid colloidal spheres, unsurprisingly, the ratio of  particle viscosity to the suspending fluid viscosity, $\lambda = \eta_o/\eta_i$, will tend towards  infinity. On the other hand emulsions are much ``softer" with viscosity ratios $\lambda$ much closer to 1. This allows emulsion droplets to function as exciting new colloidal model systems. For example, emulsion droplets have negligible friction at their surfaces with intriguing consequences for the jamming-glass crossover \cite{charbonneau2014,liu2010}, alternatively, the ease of adsorption of lipids or polymers onto droplets opens new and exciting directions in self-assembly\cite{mason1995prlb,feng2013,vandermeulen2013,desmond2013,zhang2015}.

Emulsions, with their multitude of food, care product, and petrochemical applications \cite{wibowo2001,ubbink2012}, are a system of great scientific and industrial interest, and so it should come as no surprise that the formation of emulsions is a well explored field with a huge range of emulsification techniques available. These methods include such techniques as mechanical milling, blending, high pressure homogenization, or shear mixing. While powerful, these techniques tend to produce large emulsion droplets, or emulsions with a broad size distribution. Microemulsions on the other hand, while extremely versatile due to their self assembly behaviour, have sizes in the range of 10-100 nm, and are thereby far too small for real space imaging and coordinate tracking \cite{eastoe2013}. An emulsion that combines the best of both worlds possessing both an average droplet size in the range of a few microns, with a narrow size distribution, would open up a wide range of new possibilities.

An alternative to the aforementioned methods 
 is microchannel emulsification, a versatile technique where two or more distinct fluid phases are flowed through a microscale channel and interfacial stresses are induced to generate droplets with very narrow size distributions \cite{Stone2001}. Currently microfluidic emulsification techniques remain focussed primarily on the formation of water-in-oil emulsions, usually with the \paddyspeaks{viscosity ratio}
 $\lambda$ below 1.

In order to develop the desired colloidal model system we must first consider the behaviour of a droplet in solution using the Peclet number $Pe=\tau_B/\tau_\mathrm{sed}$. Here this dimensionless parameter is the ratio between the time it takes for a particle to diffuse its own radius, $\tau_B=\sigma^3\pi\eta/8k_BT$, where $\sigma$ is the particle diameter $\eta$ is the solution viscosity and $T$ is the temperature, and the time the particle takes to sediment by its diameter,$\tau_\mathrm{sed}=\sigma/v_\mathrm{sed}$, where $v_\mathrm{sed}$ is the sedimentation velocity $v_\mathrm{sed}=\delta mg/3\pi\eta\sigma_H$, where $\delta m$ is the bouyant mass and $\sigma_H$ the hydrodynamic diameter. A Peclet number of around one is generally seen as the dividing line between a colloidal system and an athermal granular system \cite{ivlev}. As the Peclet number scales with both particle diameter and bouyant mass a colloidal emulsion can be achieved either by producing droplets of a suffciently small diameter or by using liquids with suitably similar densities \cite{wysocki2008}. As such, it is vital that a microfluidic system for the production of colloidal oil-in-water emulsions is capable of both producing droplets which are suitably small and uniform, but also to be solvent-compatible enough to produce emulsions from a range of component materials.

Microfluidic droplet generation can be realised with three types of device: T-Junction devices where a slug of the dispersed phase is extruded into a flowing fluid until the dispersed phase fluid breaks up into droplets \cite{Thorsen2001}, co-flowing devices where an outer continous phase fluid flows parallel to and surrounding an inner dispersed phase fluid until droplet generation occurs via stretching of the interface between the two fluids \cite{ganancalvo2001}, and flow-focusing devices, where co-flowing streams are forced through a narrow aperture causing droplet breakup \cite{anna2003}. Here we consider flow-focusing devices, as these provide the best system for producing droplets with sizes smaller than those of the channel dimensions.

However, producing suitable oil-in-water emulsions via microfluidic methods remains challenging. The vast majority of microfluidic emulsification systems are manufactured using polydimethylsiloxane (PDMS). While PDMS is an excellent material for water-in-oil microfluidics \cite{whitesides2002} offering good solvent compatibility, good optical transmittivity, high flexibility and great durability, it is also strongly hydrophobic, preventing the stable formation of oil-in-water droplets \cite{Makamba2003}. Although techniques for making PDMS devices hydrophilic and thereby allowing stable droplet generation exist, these are either short lived such as plasma treatment \cite{Bodas2007} or involve sequential surface coatings \cite{Bauer2010} which are impractical for use with devices small enough to easily generate micron scale droplets.

Norland optical adhesives provide a viable alternative to PDMS for oil-in-water microfluidics \cite{wagli2011, kim2007} allowing rapid templating of complex microchannels \cite{bartolo2008}. Their rapid curing speed, alongside their durability, makes them ideal for pattern transfer from PDMS molds. Additionally, while natively more hydrophilic than PDMS, exposure to an Oxygen plasma forms a long-lasting hydrophilic surface, ideal for the formation of oil-in-water droplets \cite{wagli2011}. Here we \paddyspeaks{introduce }
a microfluidic flow focussing system based on the use of NOA microfluidics for the generation of monodisperse oil in water emulsions which can produce droplets on the micron scale with polydispersities of below 5\%. To demonstrate the utility of this technique, we use the topological cluster classification to investigate the higher order structure of a polydimethylsiloxane oil emulsion, and compare this to a simulated hard sphere system.

This paper is organised as follows. In section \ref{MaterialsMethods} first we describe the micro-fabrication and polymer casting process of the Norland flow-focusing devices used. In the same section we briefly describe the flow focusing process used to generate micron scale emulsions and the confocal microscopy and coordinate tracking techniques. The topological cluster classification algorithm is used to analyse the structure of the resulting packings \cite{malins2013jcp}. In section \ref{ResultsDiscussion} we describe the droplets produced and demonstrate the excellent size selectivity of this method by varying flow rates at various viscosity ratios. We then describe the structural insights obtained from using the topological cluster classification algorithm for probing the structure of an emulsion, and our results are compared to a simulated hard sphere system. Finally, we briefly summarise our work and the scientific implications thereof in section \ref{Conclusion}.

\section{Materials and Methods} \label{MaterialsMethods}
\subsection{Imaging}
Microfluidic droplet formation was imaged using a Leica DMI 3000B Brightfield microscope. High speed droplet production was imaged using exposure times of 250 $\mu s$. Confocal microscopy as per established methods\cite{vanblaaderen1995, gasser2001,ivlev} was employed to obtain 3-dimensional images of an emulsion filled glass capillary. The imaged region consisted of $N\approx 3 \cdot 10^3$ polydimetylsiloxane (PDMS) droplets within an area of cross section 129 $\mu m$ by 129 $\mu m$ with a vertical height of 60 $\mu m$. The entire region was imaged with a LEICA SP8 confocal microscope with an excitation wavelength of 485 nm.

\subsection{Microfluidic device fabrication}

\begin{figure}[]
\centering
\includegraphics[width=\columnwidth]{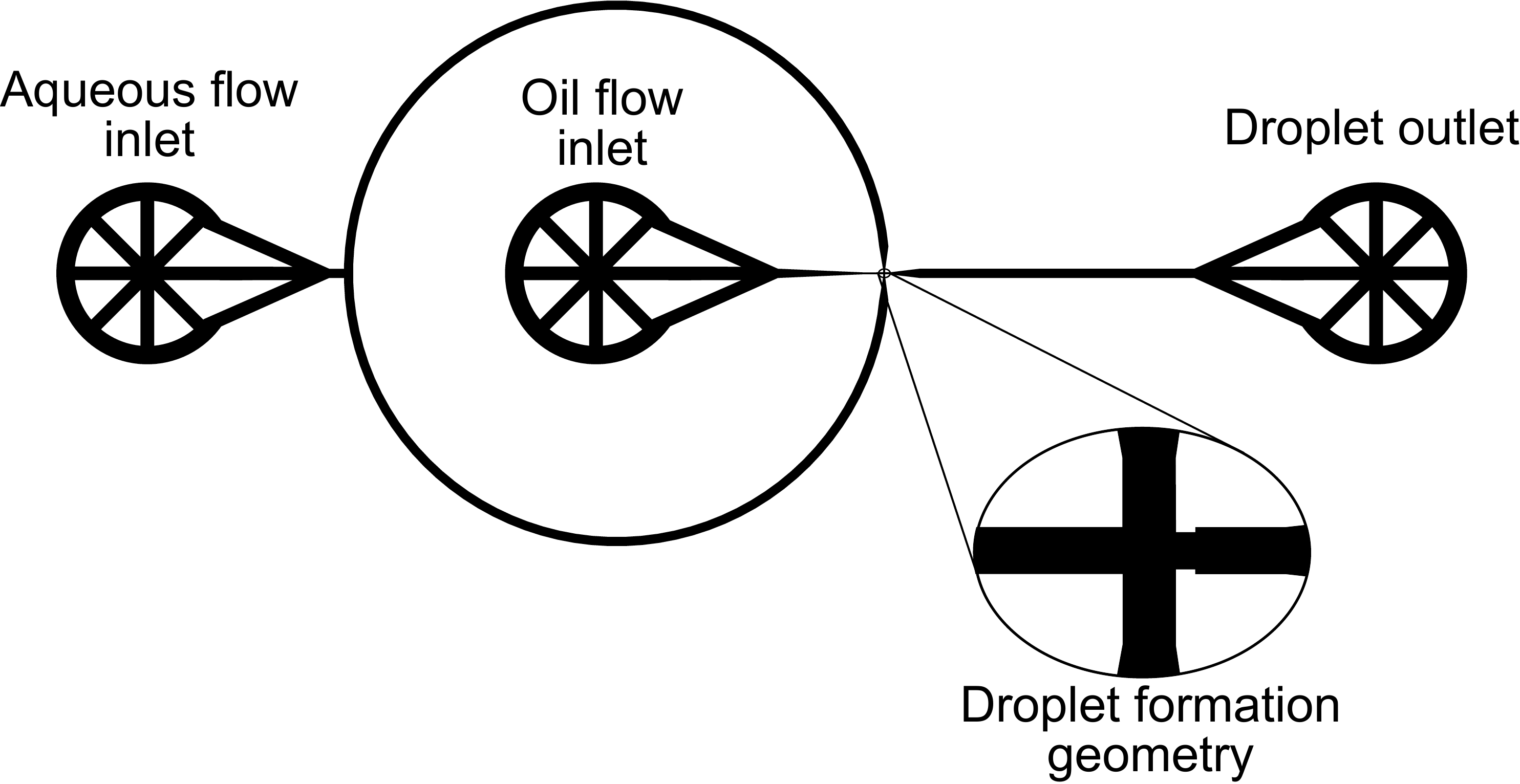}
\caption{Schematic of device layout. Popout image shows droplet formation geometry, channels measure 50 $\mu m$ before narrowing down to a 8 $\mu m$ aperture, inducing shear stress and leading to droplet formation.}
\label{figLayout}
\end{figure}

Device patterns shown in Fig.\ref{figLayout} were etched on to silicon wafers using standard photolithographic methods. Etched wafers were treated with Trichloro(1H,1H,2H,2H-perfluorooctyl) silane to allow easy siloxane lift-off. Sylgard 184 Silicone elastomer was poured on the silicon wafers, degassed under vacuum, and heat cured at 60 $^{\circ}$C for 6 hours. PDMS layers were removed from the silicon wafer, thoroughly cleaned with deionised water and ethanol. Norland optical adhesive (NOA) 81 was poured onto the prepared PDMS pattern and cured under 365 nm UV radiation for 30minutes. Tubing connections were drilled into the NOA chips, and the drilled chips were cleaned with ethanol and deionised water. NOA channels were made hydrophilic by exposing the  device layer to a 100 w Oxygen plasma for 60 seconds, and he channel geometry was enclosed by sealing the NOA chip with a thin layer of NOA produced by pressing several droplets of NOA between two clean sheets of PDMS and UV curing for 2 minutes. Once enclosed, the microfluidic channels were heat treated while for 1 hour at 60 $^{\circ}$C, before a final treatment for 2 hours using 365 nm UV radiation.

\subsection{Droplet formation}

Flow rate tests were carried out by connecting the assembled NOA chip to a nitrogen driven pressure pump (Fluigent MFCS) with polyethylene tubing. 16 mMol sodium dodecyl sulphate(SDS) aqueous solution was used as the continuous phase, and pressure and flow rate were controlled using MaesFlo 3.2 software (Fluigent, Paris, France). Flow rates were adjusted until the oil flow was stable and producing droplets. Once stable droplet formation was achieved, flow rate ratios were adjusted throughout the experimental range. Viscosity ratio measurements were carried out by varying both the oil phase used, as well as the viscosity of the aqueous phase. The liquid combinations used were an 80/20 cis-decalin/Cyclohexyl Bromide (CHB) mix with 16 mMol SDS in deionised water, an 80/20 cis-decalin/CHB mix with 16 mMol SDS in a 1:1 mix of deionised water and glycerol, and dodecane with 16 mMol SDS deionised water.

\begin{table*}[t]
\centering
\small
  \caption{\ Liquid phases used to produce the emulsion systems considered}
  \label{tbl:example}
  \begin{tabular*}{\textwidth}{@{\extracolsep{\fill}}lll}
    \hline
   Oil phase & Aqueous Phase & $\eta_o/\eta_i$ \\
    \hline
80:20 Cis-decalin/Bromocyclohexane & Deionised water & 0.3 \\
80:20 Cis-decalin/Bromocyclohexane & 50:50 Deionised water/Glycerol & 1.5 \\
Dodecane & 50:50 DIW/Glycerol & 3.8 \\
    \hline
  \end{tabular*}
\end{table*}

Droplets were subsequently collected by allowing stable droplet collection to proceed long enough to drive the produced emulsion into a 1 ml Eppendorf vial, transferred into a glass capillary and imaged on a leica brightfield microscope. Droplet sizing was carried out as shown in Fig.\ref{figDroplets}(a) where the size was taken directly from the image.

For the purpose of analysis using the topological cluster classification (TCC) analysis, droplets were produced as above using a system of 5cP Polydimethylsiloxane oil dyed with nile red as the oil phase and 32 mMol sodium dodecyl sulfate solution as the aqueous phase. Produced droplets were then collected and diluted with 50\% by weight glycerol in order to match refractive indices between the aqueous and non-aqueous phases. The index matched emulsion was filled into a capillary and imaged using confocal laser scanning microscopy on a Leica confocal microscope. A $512 \times 512 \times 256$ pixel image was taken and coordinate tracking was carried out using the Colloids particle tracking package \cite{leocmach2013}, this tracking method identifies particle centres by searching for a Gausssian of the image intensity and produces XYZ coordinates for each detected particle. XYZ coordinates of particle centers were used to calculate a 3D $g(r)$ and the Topological Cluster Classification (TCC) was used to interpret cluster populations\cite{royall2014arxiv}.

\subsection{Topological cluster classification}

The Topological cluster classification (TCC) algorithm works by searching for a configuration of particles with a bond network similar to that found in a corresponding minimum free energy cluster of up to 15 particles in a bulk assembly of spheres \cite{malins2013tcc}. By indentifying clusters related to specific interaction potentials, the TCC provides a direct link between the interactions in the system and any structural ordering found. The algorithm functions as follows: first the neighbors of each particle are identified, then the network of particles and neighbors is searched for rings of 3, 4, and 5 particles. From these rings a set of "basic-clusters" is identified, and distinguished by the number of particles that are common neighbors. Larger clusters are then identified by combining these basic clusters yielding structures with bond networks similar to the minimum energy clusters as pictured in Fig.\ref{figCluster}(c)\cite{malins2013jcp}. The emulsion droplet coordinates were analysed with the topological cluster classification. The resulting higher-order structure 
obtained in this way is compared against molecular dynamics simulations carried out in DynamO \cite{bannerman2011} using a five-component weakly polydisperse hard sphere system with previously established simulation parameters \cite{royall2015}.

\section{Results and discussion} \label{ResultsDiscussion}

Droplet production was observed and investigated under a range of flow conditions, and a strong size scaling between droplet size and flow rate ratio was observed. This relationship held true for several different solvent systems of varying viscosity ratio $\lambda$. Droplet radius and flow rate ratio were compared and two separate droplet production modes, the geometry controlled and the dripping modes, were observed. By tuning the relative flow rates droplets of a desired size and quality were then collected. To demonstrate the utility of our emulsion system as a colloidal model system, coordinate tracking and topological cluster classification analysis was carried out on \paddyspeaks{on it.}
This analysis showed a large fraction of the emulsion droplets were identified in clusters of 11 and 12 particles. This indicates  
that the emulsion has a higher order structure distinct from structure comparable to a supercooled liquid of hard spheres whose higher order structure is comprised of 10 membered defective icosahedra \cite{royall2015,royall2014arxiv,royall2016sub} \paddyspeaks{in marked contrast to the similar pair correlation at the same volume fraction.}

\subsection{Droplet production}

Droplet formation proceed in a stable manner between $Q_o/Q_i$ values of 1 to 60, outside of this range droplet formation is inhibited either by dewetting of the outer phase, or by channel filling of the inner phase.

\begin{figure}[b]
\centering
\includegraphics[width=.9\linewidth]{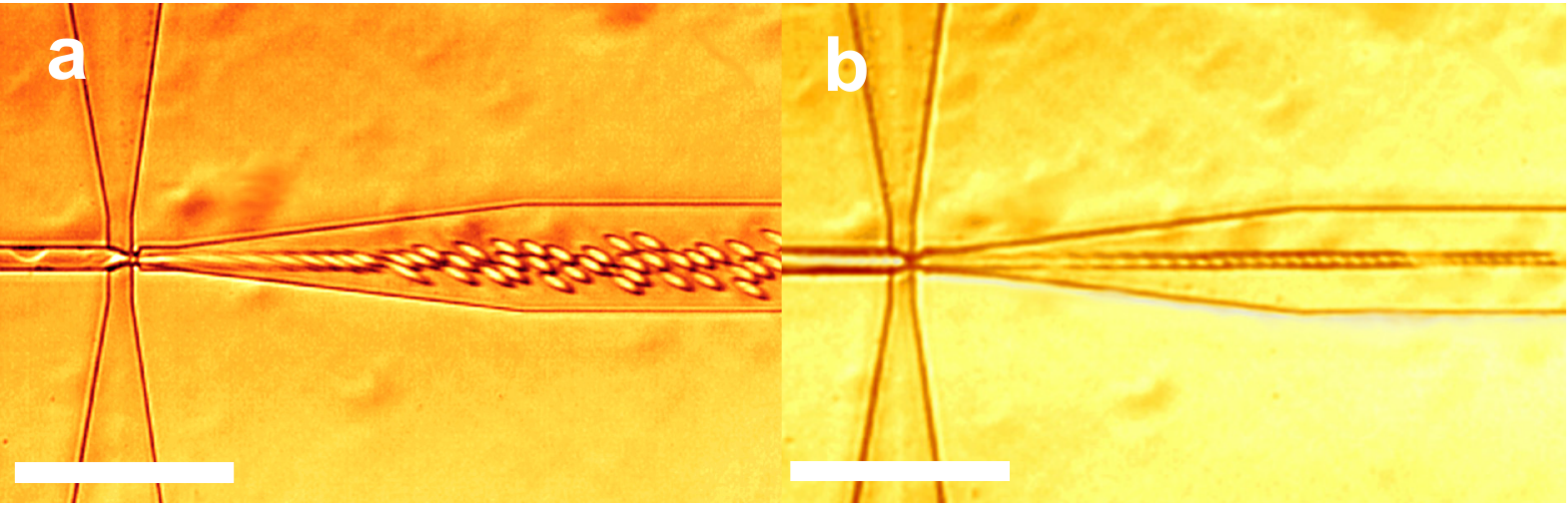}
\caption{(a) Image of geometry controlled droplet production with droplets roughly 40\% larger than channel dimensions (b) image of of droplet production via a different mode of droplet production where droplet snap-off occurs within the confines of the geometry producing droplets roughly 10\% smaller than channel dimensions. Scale bars represent 100 $\mu m$}
\label{figDropForm}
\end{figure}

Within this range of $Q_o/Q_i$ the formation of droplets was observed via two distinct mechanisms. Droplet formation in a microfluidic device is governed by three main parameters, the ratio of flow rate of the outer fluid to the inner fluid $Q_o/Q_i$, the ratio of the outer and inner fluid viscosities $\lambda = \eta_o/\eta_i$, and the capillary number $Ca=\eta_oGa_0/\gamma$ where $G$ is a characteristic deformation rate and $a_o$ is the characteristic droplet radius. Of particular importance here is the capillary number, a parameter influenced by the geometry of the device as well as the flow rate of the fluid being considered\cite{Anna2016}.
Depending on the capillary number and thereby the flow rate ratio $Q_o/Q_i$, it is possible to select between two separate modes of droplet production. At lower capillary numbers, of order $10^{-2}$, geometry controlled droplet formation occurs as shown in Fig. \ref{figDropForm}(a). In this mode the fluid interface stretches through the aperture and obstructs, leading to a pressure spike followed by droplet breakup. For this mode a weak scaling with flow rate is expected and droplet size is roughly the same as the channel geometry \cite{garstecki2005}. At higher values of the flow rate ratio $Q_o/Q_i$, with capillary number of order $10^{-1}$ this geometry controlled mode gives way to the dripping mode of droplet formation as shown in Fig. \ref{figDropForm}(b). The fluid interface in a dripping microfluidic device narrows into a fluid tendril, with the droplet ``snap-off'' point remaining stationary and within the constricting aperture. This mode of formation yields droplets smaller than the device geometry \cite{anna2006}.

\begin{figure}[b]
	\centering
	\includegraphics[width=\linewidth]{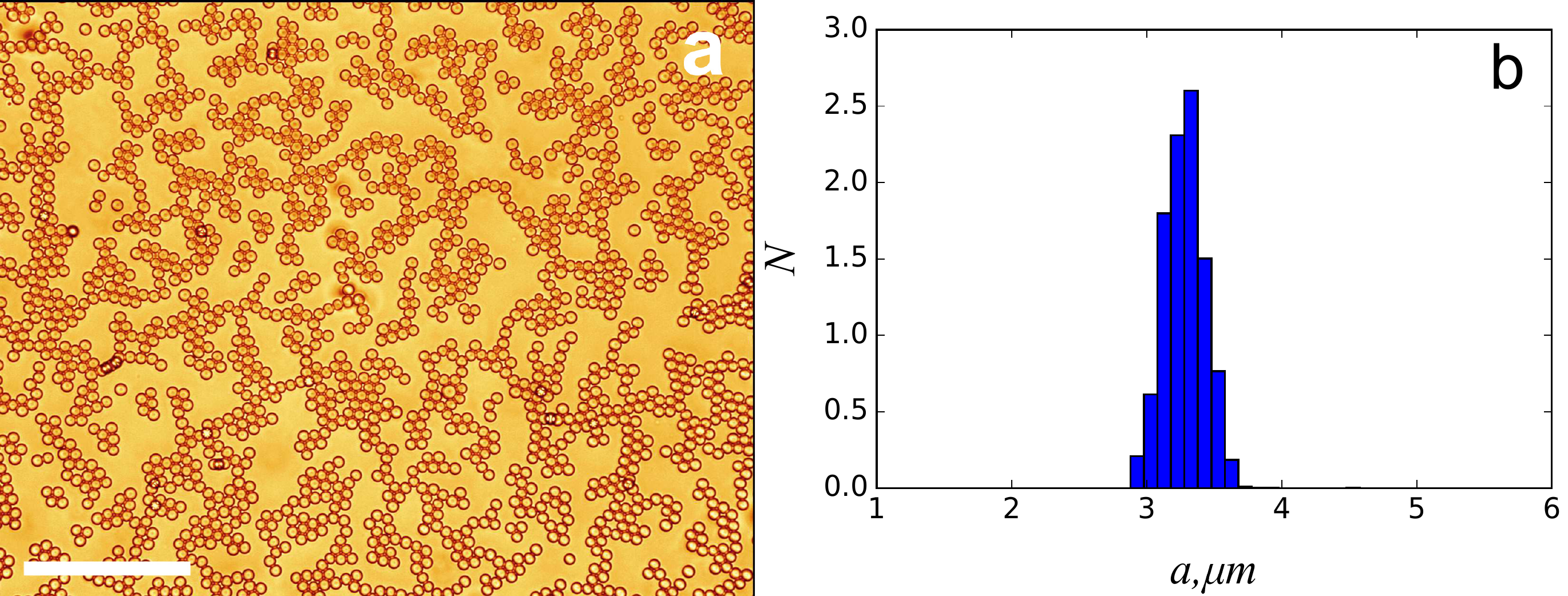}
	\caption{(a) Dodecane droplets with mean radius, $a$ of 3.3 $\mu m$ and polydispersity of 4.8\% produced in a Noa 81 microfluidic device. Scale bar represents 100 $\mu m$ (b) Normalized histogram of droplet size distribution.}
	\label{figDroplets}
\end{figure}

Once a stable formation regime was established, an emulsion produced at a fixed $Q_o/Q_i$ value of 40 was collected on the millilitre scale. Droplets produced in this way are shown in Fig.\ref{figDroplets} with a polydispersity, here defined as the relative standard deviation, of 4.8\%.

\begin{figure*}[t]
\centering
\includegraphics[width=\linewidth]{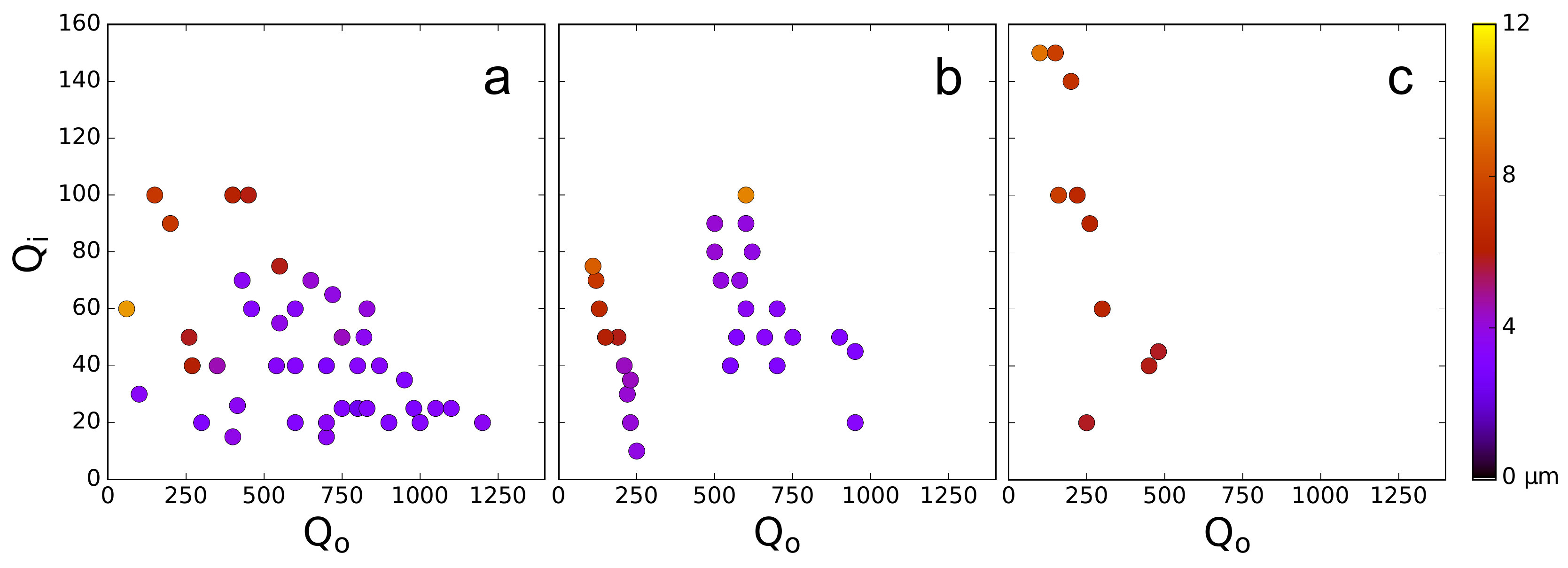}
\caption{Heatmaps of droplet radius against continuous and discrete phase flow rates for 3 different values of the viscosity ratio $\lambda$, (a) 0.3, (b) 1.5, and (c) 3.8. Flow rates $Q_o$, and $Q_i$ are shown here in units of $nl/min$. At the lower values of $\lambda$ we see a much larger droplet production range, with accessible droplet sizes ranging from 2 $\mu m$ to 12 $\mu m$. As the viscosity ratio is increased, the flow stability decreases and with it the range of accessible droplet sizes.}
\label{figHeatmap}
\end{figure*}

Varying the viscosity ratio of the outer to inner phase, $\eta_o/\eta_i$, lead to some striking differences in droplet production. At a low value $\eta_o/\eta_i$ of 0.3, droplet production was stable throughout the largest range of flow rates, with droplet size scaling inversely with flow rate as shown in Fig. \ref{figHeatmap}(a). At value of $\eta_o/\eta_i$ of 1.5, stable droplet formation was found to occur in a much narrower band of $Q_o/Q_i$, in particular droplet formation also seemed to proceed via two methods depending on the total net flow rate as shown in Fig. \ref{figHeatmap}(b) this striking difference between two methods of flow seems to suggest a switch between two different modes of droplet behaviour. Finally, at a value of $\eta_o/\eta_i$ of 3.8, droplet formation was highly unstable, and droplets produced were twice as large as at lower flow rates, with average radii above 6 $\mu m$. Surprisingly we find that droplet formation proceeds in a more stable manner at lower values of $Q_o/Q_i$, with only a small trade off in droplet size.

\begin{figure}[b]
\centering
\includegraphics[width=0.9\columnwidth]{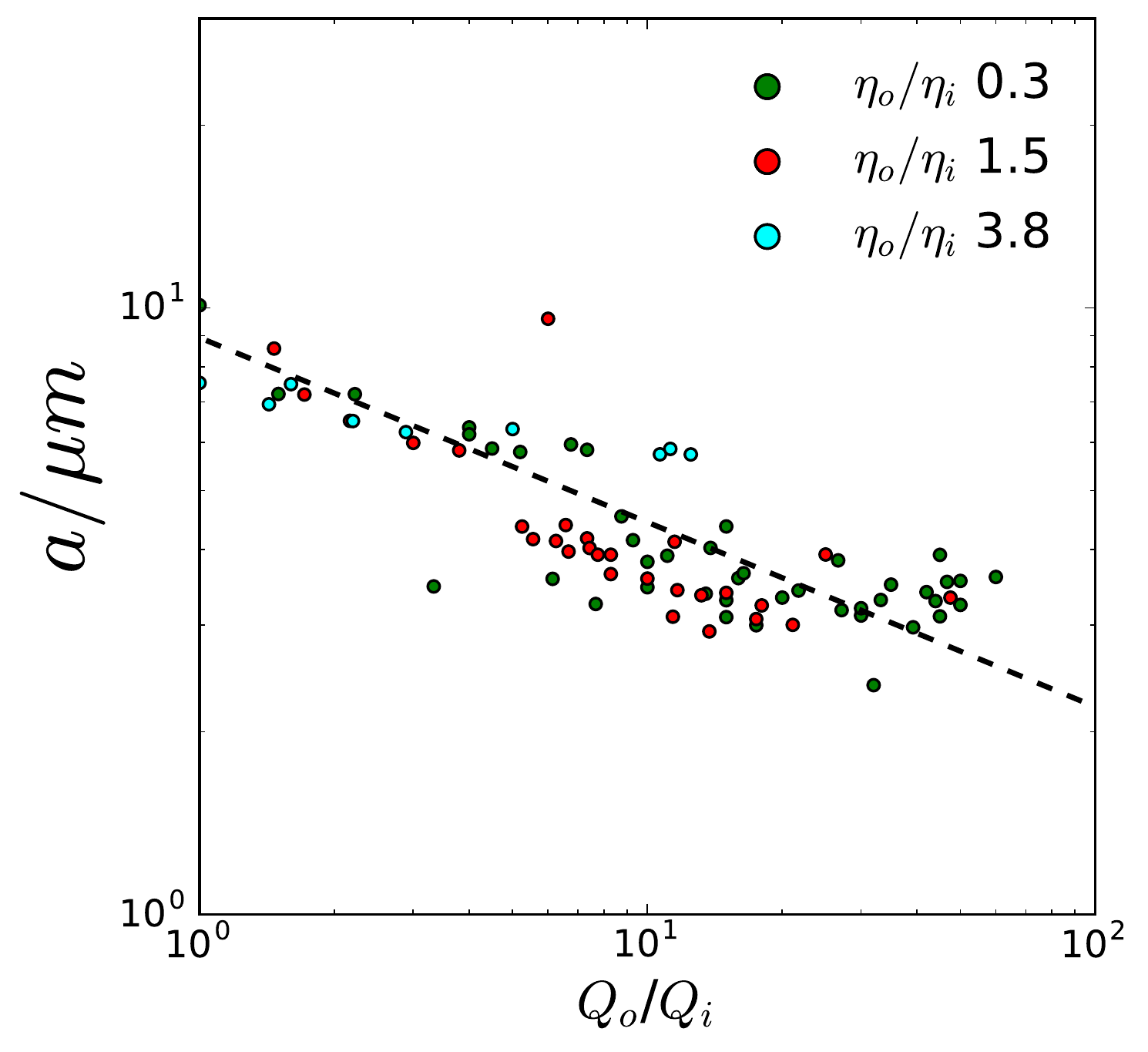}
\caption{Log-Log plot with a line to guide the eye showing scaling of droplet size with flow rate ratio \maxspeaks{corresponding to a slope of -0.268 $\mu m^{-1}$}. In particular this scaling behaves similarly \maxspeaks{across all three of the viscosity ratios examined.}}
\label{figViscRatio}
\end{figure}

As shown in Figs. \ref{figHeatmap} and \ref{figViscRatio} our microfluidic system demonstrates a high degree of versatility in the size of droplets produced. By varying $Q_o/Q_i$, a wide range of droplet sizes was produced with excellent size selectivity. Droplets produced ranged in radius from 2 $\mu m$ to 12 $\mu m$ and  droplet size was observed to be in good agreement with predictions by Garstecki \emph{et. al.} \cite{garstecki2005} as shown by the linear scaling of the log-log plot of droplet diameter, $\sigma$ and the flow rate ratio $Q_o/Q_i$ in Fig. \ref{figViscRatio} suggesting a dripping mode of droplet production.

\subsection{Emulsion structure}

To characterise the structure of the emulsion we begin by comparing the pair correlation function $g(r)$ of the emulsion system with hard spheres at the same volume fraction, $\phi=0.42$. The hard sphere system has a polydispersity of 5\%. We see that both $g(r)$ are rather similar, indicating that, at the level of two-point correlations, the emulsions are rather similar to hard spheres.

\begin{figure*}[th!]
\centering
\includegraphics[width=0.9\linewidth]{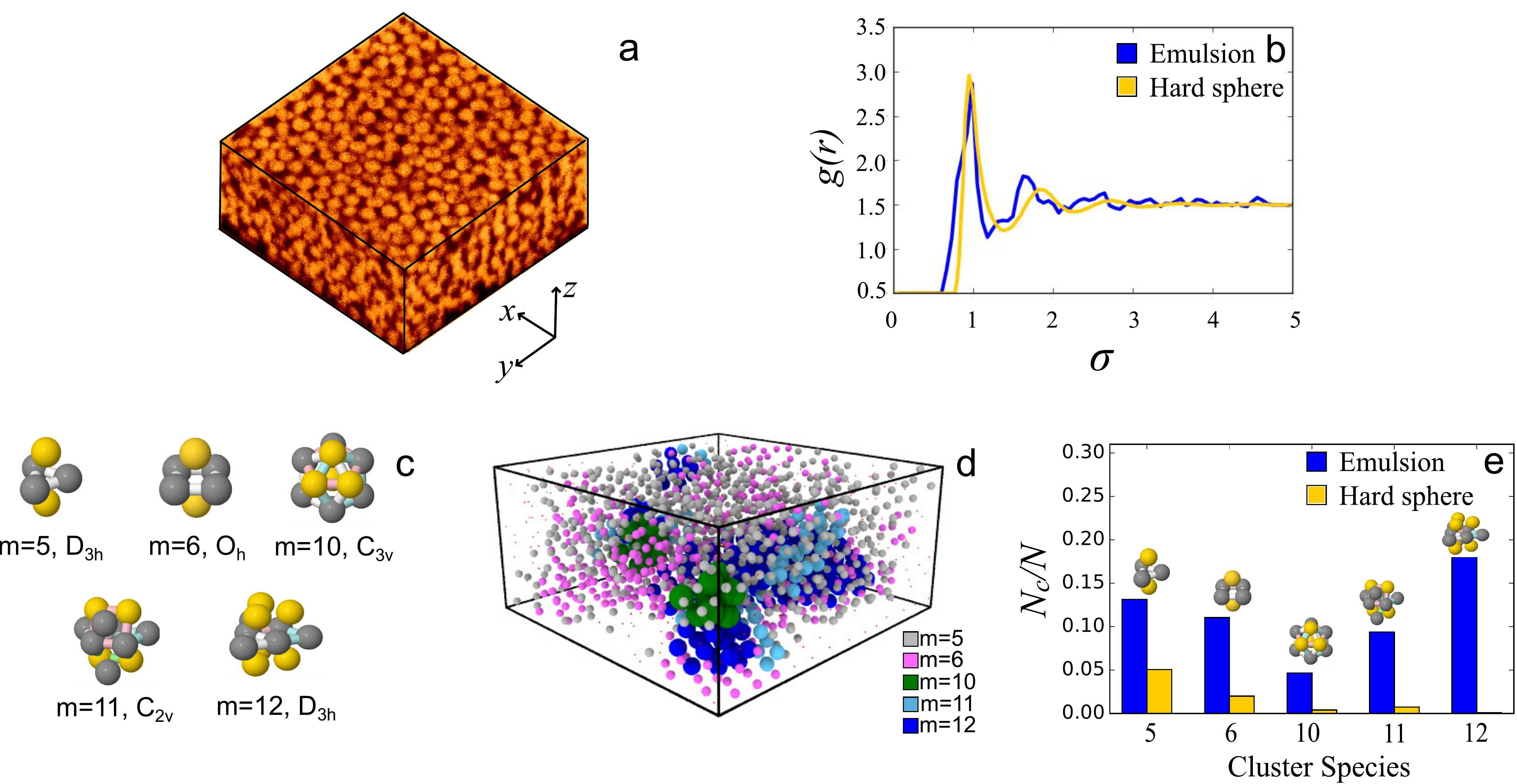}
\caption{Structural analysis of a rather monodisperse colloidal emulsion system. 
(a) Image showing representation of 3 dimensional confocal image of silicone oil emulsion with dimensions of $128.8\times128.8\times60.1$ $\mu m$, 
(b) radial distribution function of emulsion droplets in blue, and a hard sphere system of comparable volume fraction ($\phi=0.42$) in yellow,
(c) Image showing studied clusters of interest alongside respective point group symmetries, 
(d) Visualisation of cluster distribution on tracked coordinates, and
(e) Topological cluster classification chart, showing relative population of locally favoured icosahedral cluster species for emulsion in blue and comparable hard sphere system in yellow.
}
\label{figCluster}
\end{figure*}

In the case of higher-order structural measures, beyond the two-point correlations, in a dense assembly of spheres tetrahedra and consequently five membered clusters consisting of two tetrahedra are rapidly formed, with particles arranged as members of these locally favoured structures having lower potential energies and slower dynamics \cite{royall2015,royall2016sub}. \paddyspeaks{The clusters prevalent in our system are shown in Fig. \ref{figCluster}(c). In Fig. \ref{figCluster}(d) we render droplet coordinates coloured according to the identities of the clusters in which they are found.} Organisation of these tetrahedra into larger clusters can be suppressed due to dynamical arrest with the system unable to reach an equilibrated configuration for which the formation of larger clusters is expected. Due to this gels and glasses, as well as athermal granular systems 
\paddyspeaks{are }
dominated by tetrahedra, while dense systems which are thermal, and in which reorganisation can occur such as (supercooled) liquids) would be dominated by higher membered clusters \cite{royall2008,taffs2013}. Here the hard spheres are $\phi=0.42$, which isn't \paddyspeaks{is not sufficiently dense to be supercooled}
and in fact the most prevalent structures found by the TCC are smaller $m=5,6$ clusters [Fig.\ref{figCluster}(e), yellow]. On the other hand, in our emulsion system we
see significantly more $m=12$ and $m=11$ structures, where $m$ is the number of particles in the structure, than $m=5$ structures [Fig.\ref{figCluster}(e), blue]. 
Such larger membered clusters suggest that our system is a thermal system analogous to a supercooled liquid \cite{royall2015}. 
This moreover shows that the higher order structure is rather distinct from that of supercooled liquids hard spheres (formed at higher $\phi\gtrsim0.56$) which are dominated by 10-membered defective icosahedra \cite{royall2015,royall2014arxiv,royall2016sub}. This suggests that the emulsion system exhibits an interaction potential very distinct to that of hard spheres, perhaps due to electrostatic effects, or the ease with which the droplets can be deformed.

\section{Conclusions} \label{Conclusion}
In this work we have demonstrated a powerful and versatile microfluidic droplet generator for the production of oil-in-water emulsions, and shown that this system produces droplets of a wide range of sizes with \paddyspeaks{very }
low polydispersity. By utilising the favourable mechanical and surface chemical properties of Norland optical adhesives we open up a new and promising parameter space for studying colloidal systems with carefully tuned chemistry and viscosity ratios. We demonstrate the potential of this system as a colloidal model system by studying the arrangement of emulsion droplets with respect to locally favoured structures and as shown in Fig. \ref{figCluster}(e) are able to obtain detailed information as to the local structure of a highly uniform emulsion. By comparing the population of $m=11,12$ clusters with respect to the the hard spheres which at this volume fraction are dominated by small $m=5,6$ clusters, there is much more higher-order structure in the emulsion system. Moreover this higher order structure is distinct from the $m=10$ defective icosahedral structure formed in hard spheres even at higher density. Thus the 
the silicone oil emulsion is not \paddyspeaks{dynamically} arrested and capable of rearrangement, behaving much like a supercooled with well-developed higher-order structure which is yet distinct from hard sphere supercooled liquids.
We believe that this microfluidic system shows great potential as a powerful tool for studying exciting new colloidal behaviour, and in particular by carefully turning the microfluidic flow properties and the relative fluid densities the technique demonstrated here can allow the production of truly colloidal emulsions with tunable size, size distribution, and composition.

\section{Acknowledgements}
The authors would like to thank Denis Bartolo, Oliver Dauchot, and Francesco Turci for many helpful discussions. M.M. acknowledges the EPSRC Doctoral training allowance EP/L504919/1, CPR acknowledges the Royal Society and Kyoto University SPIRITS fund for financial support, JD and CPR acknowledge the European Research Council (ERC consolidator grant NANOPRS, project number 617266), and JD acknowledges Bayer CropScience AG for financial support.

\bibliographystyle{unsrt}
\bibliography{newModel}

\end{document}